\documentclass[prl,twocolumn,showpacs,floatfix]{revtex4}
\usepackage{amsfonts}

\usepackage{amssymb}
\usepackage{amsmath}
\usepackage{graphicx}

\begin{document}

\title{Master Equation and Control of an Open Quantum System with Leakage}
\author{Lian-Ao Wu$^{1,3}$, Gershon Kurizki$^2$ and Paul Brumer$^1$}
\affiliation{$^1$Center for Quantum Information and Quantum Control and \\
Chemical Physics Theory Group, Department of Chemistry, University of
Toronto, \\
Toronto, Ontario M5S 3H6, Canada\\
$^2$ Chemical Physics Department, Weizmann Institute of Science, Rehovot,
Israel\\$^3$
Department of Theoretical Physics and History of Science, The Basque 
Country University (EHU/UPV), PO Box 644, 48080 Bilbao, Spain}

\begin{abstract}
Given a multilevel system coupled to a bath, we use a Feshbach
$P,Q$ partitioning technique to
derive an exact trace-nonpreserving master equation for a
subspace $\mathcal{S}_{i}$ of the system. The resultant equation
properly treats the leakage effect
from $\mathcal{S}_{i}$ into the remainder of the
system space.
Focusing on a second-order approximation, we
show that a one-dimensional master equation is sufficient to
study problems of quantum state storage and is
a good approximation, or  exact,  for several analytical
models. It allows a
natural definition of a leakage function and its control, and provides a
general approach to study and control decoherence and leakage.
Numerical calculations  on an harmonic oscillator coupled to a room temperature
harmonic bath show that the leakage can be suppressed by the
pulse control technique without requiring ideal pulses.
\end{abstract}

\pacs{3.65.Yz, 03.67.Pp, 37.10.Jk}
\maketitle

\textit{Introduction.---} Control of quantum dynamics is of great
interest for ``quantum technology industries", such
as quantum computing. The control of closed quantum systems is well
established, and has been
extensively studied in areas such as chemical physics \cite%
{Brumer02,Rice}. Efforts to extend these studies to open
systems, where the system interacts with an environment, are now
underway\cite{WuBrumer}. Quantum information processing has already expended
considerable effort on open systems. Hence, we anticipate that
methods being developed in the latter area may well be useful in
the former, and vice-versa \cite{Betal}.

A fundamentally difficult problem in quantum information
processing is that of decoherence, i.e., the loss of quantum
information in a system due to its interaction with its
environment (or ``bath") \cite{Nielsen00, Kempe01}. For
multi-level systems, such as molecules, the interaction can also
cause leakage (i.e. loss of population) from a system subspace of
interest, denoted $\mathcal{S}_i$, into the system space outside
of $\mathcal{S}_i$. Theoretical strategies for combating such
deleterious environmental effects in the absence of natural
decoherence free subspaces \cite{Kempe01,Zanardi97},   invoke the
dynamical control of system-environment interactions by external
fields \cite{Viola98, Agarwal99, Wu020, Lloyd00, Lidar05,
Kurizki04, Alicki-etal}.

The aim of dynamical control in open systems is to suppress
effects of the environment in order to control system processes at
will. For example, ideal Bang-Bang (BB) control  of decoherence,
decay and leakage \cite{Wu020, Lloyd00} utilizes idealized
zero-width pulses and the Trotter formula to achieve this goal,
although higher-order finite-pulse widths have been considered in
model cases \cite{Lidar05}.

Proposals to control decoherence by realistic (non-ideal) pulses
have often invoked Zwanzig's projection operator techniques
\cite{Kurizki04, Alicki-etal} resulting in a differential master
equation for the density operator of the system up to second order
in the system bath coupling. This equation allows the tractable
treatment of complicated quantum dynamical processes, eliminating
the ideal zero width pulses and Trotter formula assumptions and
allowing natural dynamical evolution and dynamical control on an
equal footing.  For example, Ref. \cite{Kurizki04} provides a unified way to
suppress the decoherence of two level systems by arbitrary fields
that control the system-bath interaction.


Below, we consider an $N$ dimensional system ($N$ can be
infinite), spanned by the bases $\left\{ \left\vert n\right\rangle
\right\} $, coupled to a bath, and develop procedures to protect
quantum information stored in a $d$-dimensional subspace
$\mathcal{S}_{i}$ of the system. Specifically, 
we use a projection operator approach to obtain an exact
trace-\textit{non}preserving master equation 
for the dynamics of the system subspace $\mathcal{S}_{i}$. We then  
introduce a characteristic leakage function which is used to 
consider dynamics and control of this
subspace to second order in the system-bath interaction.

\textit{Zwanzig's Projection-Operator Approach.--- }
The most general Hamiltonian of the $N$ dimensional system  plus the bath is
\begin{equation}
H=H_{0}+H_{I} = H_S + H_B + H_I,  \label{eq1}
\end{equation}
where $H_S$ and $H_B$ are the system and bath Hamiltonians, respectively,
and the system-bath interaction is
$H_{I}=\sum_{\alpha }S_{\alpha }B_{\alpha }$, where $S_{\alpha }$ and $
B_{\alpha }$ are Hermitian system and bath operators. During the dynamics,
the information stored in the system subspace of interest $\mathcal{S}_{i}$
will distribute into the bath and \emph{leak} into the system states outside
of $\mathcal{S}_{i}$ if one does not protect the subspace. Therefore, in order
to store the information in $\mathcal{S}_{i}$, we need to control the system
through an interaction with an
external field $H_{c}(t)$, protecting the information during the course
of time $t$. The total system Hamiltonian including control is therefore
$H_{S}(t)=H_{S}+H_{c}(t)$.

We first derive a \emph{closed} master equation for $\mathcal{S}_{i}$. Since $
\mathcal{S}_{i}$ opens both to bath and to other system states outside of $
\mathcal{S}_{i}$, effects of leakage have to be considered in deriving a
master equation. As usual, the derivation begins in the interaction
representation with respect to $H_0$,
in which the equation of motion is $\frac{\partial }{
\partial t}\,\rho (t)=-i[H_{I}(t),\rho (t)]\equiv \mathcal{L}(t)\rho (t)~$.
Here the system-bath interaction is  $H_{I}(t)$ in the interaction
representation and the Liouville superoperator $\mathcal{L}(t)$ is defined by
this equation \cite{Breuer02}. The superprojection operation $\mathcal{P}$
that we seek defines the  \emph{relevant} part of the total density
matrix \cite{Breuer02} $\mathcal{P}\rho (t)$ for our new open system, i.e.
the subspace $\mathcal{S}_{i}$ of the entire system. Specifically, the
superprojection operation comprises two commuting parts: a trace over the
bath components, and a projecting out of the $\mathcal{S}_{i}$ subspace from
the full $N$-dimensional space of the system. The associated superprojector
$\mathcal{P}$ is therefore defined as
\begin{equation*}
\mathcal{P}\rho =\mathbf{P}\text{ tr}_{B}\{\rho \}\mathbf{P}
\otimes \rho _{B}\equiv \eta \otimes \rho _{B},
\end{equation*}
where $\mathbf{P}=\sum_{i=0}^{d-1}\left\vert i\right\rangle \left\langle
i\right\vert $ denotes a projection onto $\mathcal{S}_{i}$ and $\eta $ is
the relevant part of the total density matrix, which is projected from the
total system
density matrix $\rho _{S} =$ tr$_B \rho$ as $\eta \equiv \mathbf{P}
\rho _{S}\mathbf{P}$.
The matrix $\rho_B$ is chosen as the initial state of the bath.
The trace of the matrix
$\eta $ is not necessarily one, but $\eta $ satisfies a \emph{closed} equation as
does $\rho _{S}$, and plays the same rule as $\rho _{S}$, but in $\mathcal{S}_{i}$.
That is, an arbitrary system observable acting only on $\mathcal{S}_{i}$
obeys the relation $O=\mathbf{P}O\mathbf{P}$. The expectation value of
the operator $O$ when the system+bath is in state $\rho$ is
tr$\{\rho O\}=$tr$\{\mathbf{P}\rho _{S}\mathbf{P}O\}=$tr$\{\eta O\}$, which
is the same as the
expectation value of the operator $O$ in the state characterized only by $
\eta $. This implies that the matrix $\eta $ provides a complete description of
the physics in $\mathcal{S}_{i}$, in the same sense that $\rho _{S}$
completely describes the total open system.

Applying the superprojector to the equation of motion for $\rho$
gives a time-local master equation
\begin{equation}
\frac{\partial }{\partial t}\,\mathcal{P}\rho (t)=\mathcal{K}(t)\mathcal{P}
\rho (t),  \label{eq2}
\end{equation}
where $\mathcal{K}(t)$ is the time-convolutionless generator \cite{Breuer02}.
Unlike the usual approach, our new system is a $d-$dimensional
subspace of the $N$-dimensional space of the total system. Alternatively,
Eq. (\ref{eq2}) can be derived by applying the Feshbach projection operator
approach\cite{Shore,Rice2} to the traditional trace-preserving master equation for $\rho_S$.

Equation (\ref{eq2}) is exact and holds for almost\cite{Breuer02} all
arbitrary systems and interactions, and for initial conditions $\rho (0)=
\mathcal{P}\rho (0)$, i.e. where the quantum system is initially
within $\mathcal{S}_i$. Since population can flow out of this subspace,
the master equation is not trace-preserving.
Unfortunately this equation is as difficult
to solve as the original equation. Therefore, perturbation expansions are
needed in order to apply the result to actual problems.

To second order in the coupling strength of the interaction, $\mathcal{K}
(t)=\int_{0}^{t}ds\mathcal{PL}(t)\mathcal{L}(s)\mathcal{P}$. Introducing the
explicit expressions for the projection operator and the Liouville
superoperator, we can obtain the second-order $d\times d$ dimensional master
equation in the interaction representation,
\begin{equation}
\frac{\partial }{\partial t}\,\eta (t)=-\lambda ^{2}\int_{0}^{t}ds\mathbf{P}
\text{ tr}_{B}[H_{I}(t),[H_{I}(s),\eta (t)\otimes \rho _{B}]\mathbf{P}.
\label{eq3}
\end{equation}
Here $\lambda H_{I}$ replaces $H_{I}$, with the small parameter
$\lambda $ introduced to characterize the order of perturbation expansion.
For the single component interaction $H_{I}(t)=S(t)B(t)$, Eq. (\ref{eq3}) can be
considerably simplified.

\textit{One Dimensional Dynamics and the Principle of Control.---}
A primary example is the dynamics, control and protection of
one normalized state $\left\vert \phi \right\rangle $,
in the interaction representation, within the
$d-$dimensional subspace $\mathcal{S}_{i}$. [Spontaneous emission, for
example,  is a case
where $\left\vert \phi \right\rangle$ is an energy eigenstate].
In general $ \left\vert \phi \right\rangle $ is a superposition
of eigenstates, rather than a single eigenstate, and
we can rearrange the bases of $\mathcal{S}_{i}$ so
that $\left\vert \phi \right\rangle $ is one of the new orthonormal basis 
elements.

Suppose that the initial state $\eta (0)=\left\vert \phi \right\rangle
\left\langle \phi \right\vert $. The subsystem evolves
according to the closed equation (\ref{eq3}) with $d=1$ and, at time $t$,
$\eta(t)=b(t)\left\vert \phi \right\rangle \left\langle \phi \right\vert $,
where, in general, $b(t)$ is written as
\begin{equation}
b(t)=\exp (-L(t)), \label{eq30}
\end{equation}
with $b(t)\leq 1$ or $L(t)\geq 0.$ Substituting $\eta (t)$ into the
master equation (\ref{eq3}) with $d=1$, gives an analytic expression second order
in $\lambda $,
\begin{equation}
L(t)=\lambda ^{2}\int_{0}^{t}ds C(s),
\label{eq4}
\end{equation}
where
\begin{equation}
C(s)=\int_{0}^{s}ds^{\prime
}\sum_{\alpha \beta }[\mathfrak{S}_{\alpha \beta }(s,s-s^{\prime })\Phi
_{\alpha \beta }(s^{\prime })+h.c.],  \label{eq5}
\end{equation}
and $\mathfrak{S}_{\alpha \beta }(s,s^{\prime })=\overline{\Delta
S_{\alpha }(s)\Delta S_{\beta }(s^{\prime })}$ . Here $\Delta S_{\alpha
}(t)=S_{\alpha }(t)-\overline{S_{\alpha }(t)}$ and $\overline{S_{\alpha }(t)}%
=\left\langle \phi \right\vert S_{\alpha }(t)\left\vert \phi \right\rangle $.
$\Phi _{\alpha \beta }(t-s)=$tr$_{B}[B_{\alpha }(t)B_{\beta }(s)\rho _{B}]$
is the bath correlation function for multi-term system-bath interactions.
Note that $C(s)$ is a linear function of matrix elements $\Phi_{\alpha \beta}$.
We term
the time-dependent $L(t)$ a \textit{leakage function}, by analogy with the
decoherence function \cite{Breuer02}. It describes the leakage from
$\mathcal{S}_{i}$ due to the bath and into the space outside of $\mathcal{S}_{i}$.
Higher than second order effects in the leakage function are included in
Eq.(\ref{eq2})

$L(t)$ is a functional of the initial state
$\left\vert \phi \right\rangle$ and any added control $H_{c}$, the latter 
generally through incident external fields. 
Given a time $t$, the solution
of the variational equation $\delta L(t) =0$ with respect to the
state $\left\vert \phi \right\rangle $ in the absence of $H_c$  yields, 
to second order, a self protected state. Alternatively, solution to this variational 
equation with respect to the incident electromagnetic fields for fixed
$\left\vert \phi \right\rangle $ provides optimal control fields, to
second order, to protect $\left\vert \phi \right\rangle $ against decoherence.
Later in this letter we address  optimizations with respect to 
the control fields for realistic molecular systems. Optimizations with respect
to $\left\vert \phi \right\rangle $ are currently under study.

As in all approximation techniques, the utility of the second order
approximation [Eq. (\ref{eq3})] is examined by
comparison with exact cases. We  consider two examples.

\textit{Example I: Pure leakage}.  Consider a pure
leakage case, where the system is a one-dimensional Harmonic oscillator in
which there is no system-bath interaction. The system is described by
$H_{S}=\omega a^{\dagger }a$ and is polarized by the interaction $H_{I}=\lambda
(a^{\dagger }+a)\otimes I_{B}$, where $I_{B}$ is the unit operator of a
bath and $a^{\dagger }(a)$ is a bosonic creation
(annihilation) operator for the harmonic oscillator. For the case of the
ground state $\left\vert \phi \right\rangle =\left\vert 0\right\rangle $,
the exact solution is $b_{\text{ex}}(t)=\left\langle 0\right\vert \rho
_{S}(t)\left\vert 0\right\rangle =\exp (-\frac{4\lambda ^{2}}{\omega ^{2}}
\sin ^{2}\frac{\omega t}{2})$. The second-order solution [Eqs. (\ref
{eq30}) and (\ref{eq4})] gives the same result.
When
$\left\vert \phi \right\rangle =\left\vert 1\right\rangle ,$ the exact
analytical solution $b_{\text{ex}}(t)=(1-\frac{4\lambda ^{2}}{\omega ^{2}}%
\sin ^{2}\frac{\omega t}{2})^{2}\exp (-\frac{4\lambda ^{2}}{\omega ^{2}}\sin
^{2}\frac{\omega t}{2})$, while $b(t)=\exp (-\frac{12\lambda ^{2}}{\omega
^{2}}\sin ^{2}\frac{\omega t}{2}).$ In the superposition case of $\left\vert
\phi \right\rangle =\frac{1}{\sqrt{2}}(\left\vert 0\right\rangle +\left\vert
1\right\rangle ),$ $b(t)=\exp (-\frac{4\lambda ^{2}}{\omega ^{2}}(\sin ^{2}%
\frac{\omega t}{2}+\sin ^{4}\frac{\omega t}{2}))$ and the exact solution
$b_{\text{ex}}(t)=[1-\frac{4\lambda ^{2}}{\omega ^{2}}(1-\frac{\lambda ^{2}}{%
\omega ^{2}})\sin ^{4}\frac{\omega t}{2}]\exp (-\frac{4\lambda ^{2}}{\omega
^{2}}\sin ^{2}\frac{\omega t}{2}).$ Their second-order expansions are all
the same and are essentially equal to one another for $\frac{\lambda ^{2}}{\omega
^{2}}<0.1$.

\textit{Example II: Spin-bath model within the rotating wave approximation.}
Here the system Hamiltonian reads
$H_{0}=\epsilon \sigma ^{z}+\omega a^{\dagger }a$ and $H_{I}=\lambda (\sigma
^{+}a+\sigma ^{-}a^{\dagger }),$  where $\sigma ^{\pm }=(\sigma
^{x}\pm i\sigma ^{y})/2$. Let $\left\vert \phi \right\rangle =\left\vert
0\right\rangle $ and let the bath be in the state $a^{\dagger }\left\vert
V\right\rangle ,$ where $\left\vert 0\right\rangle $ is the spin-down state
and $\left\vert V\right\rangle $ is the vacuum bath state. Solving the
problem exactly gives $b_{\text{ex}}(t)=1-\sin ^{2}(t\sqrt{(\frac{\epsilon
-\omega }{2})^{2}+\lambda ^{2}})\frac{\lambda ^{2}}{\sqrt{(\frac{\epsilon
-\omega }{2})^{2}+\lambda ^{2}}}$. The second-order solution to the master
equation is $b(t)=\exp (-\frac{16\lambda ^{2}}{(\epsilon -\omega )^{2}}\sin
^{2}\frac{(\epsilon -\omega )t}{2})$, which agrees with the exact result in
second order. They can also be shown numerically to be in good agreement
when $\frac{\lambda ^{2}}{(\epsilon -\omega )^{2}}<0.1$.

{\it Quantum Control:}
Within the framework outlined here, the goal of quantum control becomes:
given a time $t$ of interest, we computationally
seek the solution of the variational equation $\delta
L(t)=0 $ with respect to $H_{c}$, with the inclusion of any physical constraints
of the control, The result is the
control functional $H_{c}(s^{\prime }).$  Note that this approach
does not just optimize $C(s)$ itself at time $s$,  but rather includes
the history of the time evolution of $C(s).$

Although idealized BB control provides a possible mathematical
solution, of $L(t)=0$, it requires unrealistic (zero-width)
pulses. Hence, our focus is to replace idealized control by an
approximate variational or numerical solution that minimizes
$L(t)$ under realistic pulse energy and pulse width constraints.


\textit{Harmonic Oscillator coupled to a harmonic bath.-} As an
application of this framework, consider as the system the harmonic
approximation (frequency $\Omega )$ to a Morse oscillator
\cite{Wang01,Paul04} for, e.g., molecular iodine, which allows us
to employ a simplifying symmetry. The total Hamiltonian, where the
bath has $\ell$ oscillators, is
$H=\Omega a^{\dagger }a+\sum_{j=1}^{\ell}\omega _{j}a_{j}^{\dagger
}a_{j}+\lambda SB,$
where $B=\sum \alpha (\omega _{j})(a_{j}+a_{j}^{\dagger })$. The interaction
is separable with $S(t)=e^{-i\Omega t}a+e^{i\Omega t}a^{\dagger }$ and  the bath
correlation function is $\Phi (t)=\sum \alpha ^{2}(\omega _{j})\{[1+n(\omega
_{j})e^{-i\omega _{j}t}+n(\omega _{j})e^{i\omega _{j}t}\}$ where $n_{j}=%
1/[\exp (\beta \omega _{j})-1]$. At low energy,
I$_{2}$ vibrational motion is harmonic, with $\Omega
=213.7$cm$^{-1}$ and $\alpha (\omega _{j})=\sqrt{\frac{\omega _{j}\omega
_{d}(1-e^{-5})}{40\pi \ell}}$ with $\omega _{j}=-\frac{\omega _{d}}{2}\ln (1-
\frac{j(1-e^{-5})}{\ell})$ , where $\omega _{d}$ is the cut-off frequency at
$j=\ell$.

As an example, consider a superposition of the eigenstates of the system
harmonic oscillator
as the state in need of protection in a bath at room temperature.
Such states would be of interest, for example, in
pump-dump coherent control scenarios\cite{Brumer02,Rice} where this is the
initially pumped state. In that case one would be
interested in maintaining this state over time scales of $\approx 700$ fs,
the system decoherence time\cite{Paul04}.

\begin{figure}[th]
\centering {\includegraphics[angle=0,scale=0.30]{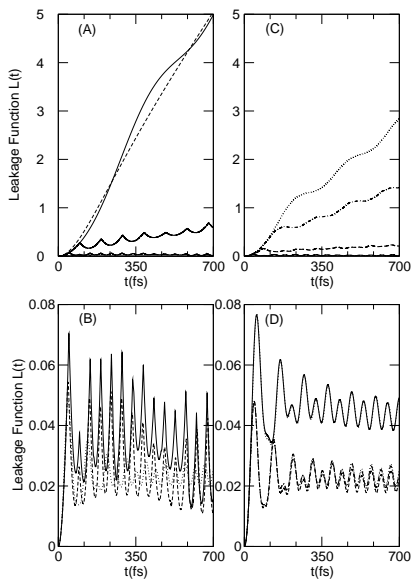}}
\caption{(A) (upper solid curve) $L(t)$ in units of $\protect
\lambda ^{2}$ for an Harmonic oscillator, in the initial state,
$\protect \left\vert \phi \right\rangle =\frac{1}{\sqrt{2}}
(\left\vert 0\right\rangle +\left\vert 1\right\rangle )$ 
coupled to a
bath of 1000 harmonic oscillators at $T=$300K. Here $\protect
\omega _{d}=0.01(fs)^{-1} \approx 1.5\Omega $.
The dashed curve shows the dominant part,
as discussed in the text. The three lower solid curves show $L(t)$
with $\Delta =0$, $\protect\phi _{0}=\protect\pi $ and
$\protect\tau =\frac{\protect\pi }{20\Omega },\frac{%
\protect\pi }{10\Omega }$ and $\frac{\protect\pi }{5\Omega }$, in the
order from bottom to top. (B): $L(t)$ with $\protect\tau =%
\frac{\protect\pi }{10\Omega }$ and $\protect\phi _{0}=\protect\pi $, but
different values of $\Delta $. Solid curve $\Delta =\protect\tau /5;$ Dashed
curve: $\Delta =\protect\tau /2;$ Dotted curve: $\Delta =\protect\tau .$
(C): $L(t)$ with $\protect\tau =\frac{\protect\pi }{10\Omega }$ and $%
\Delta =\frac{\protect\tau }{2}$ but different pulse intensities,
$\protect\phi _{0}=\frac{\protect\pi }{10},\frac{\protect\pi }{5},\frac{%
\protect\pi }{2}$ from top to bottom. (D): two lower curves, $L(t)$
with $\protect\phi _{0}=\protect\pi /100$, $\protect\tau =\protect%
\tau _{0}/100$ $(\protect\tau _{0}=\frac{\protect\phi _{0}}{10\Omega })$ \
and $\protect\phi _{0}=\protect\pi /10$ and $\protect\tau =\protect\tau %
_{0}/10$ all with $\Delta =\protect\tau /2.$ They have the same value of $%
\Omega _{c}.$ The upper curves have the same values of $\protect\phi _{0}$
and $\protect\tau $ but different signs of $\protect\phi _{0}$. }
\end{figure}

Figure 1A shows $L(t)$ for the parameters shown in the figure caption.
In some cases, $\mathfrak{S}(s,s-s^{\prime })$
may be a function of $(s-s')$ only, in which case $L(t) \propto t$, shown as
the dashed line in Fig. 1A.
This linear dependence on $t$ is similar to
that of the usual decoherence function \cite{Breuer02} in the long time
limit. However,  at shorter times the dependence of
$\mathfrak{S}(s,s-s^{\prime })$ on $s$ is seen to contribute non-negligibly, leading
to oscillatory $L(t)$.
Below we will numerically study the behavior of the leakage function in the
short-time region to determine the extent to which leakage can be controlled.

\textit{Control.-}
Physically, the origin of the control is that the frequency of the
system (here, an harmonic oscillator) is periodically,
dynamically, Stark shifted by the alternating field. To this end
we employ the realistic control Hamiltonian $H_{S}(t)=(\Omega
+f(t))a^{\dagger }a$, which results from strong laser pulses
acting on electrons that induce an additional time-dependent
nuclear potential. We model the control function as a periodic
rectangular interaction: $f(t)=0$ for regions other than $n\tau -
\Delta<t<n\tau$, $n$ integer. Inside these regions $f(t)$ is
defined so that $\phi_0 = \int_{\tau-\Delta}^{\tau}  f(t) dt$.
That is, for nonzero $\Delta$,  $f(t)=\phi _{0}/\Delta $ over the
control interval, and for $\Delta=0$, $f(t) = \delta(t-n\tau)$.
The functional form contains three main control parameters: the
time interval $\tau $, the pulse width $\Delta $ and the
interaction intensity $\phi _{0}$. For comparison with realistic
pulses, we show $L(t)$ with ideal impulsive phase modulation
($\Delta=0$) in the three lower curves in Fig. 1 (A). Clearly, the
shorter the control interval, the better the control.

Figure 1(B) shows $L(t)$ with fixed $\tau$
 and $\phi_0$, but with different pulse widths
$\Delta $. The results show that the quality of the control is only weakly
dependent on the pulse width. For example, the control is excellent even if
the width of
the pulse is equal to the control interval $\tau $. In this case the control
is equivalent to adding a constant frequency $\Omega _{c}=\phi _{0}/
\tau $ to the harmonic oscillator frequency, i.e. shifting the system
frequency by  $\Delta /\tau =1$ means shifting the system
frequency to ($\Omega +$ $%
\Omega _{c}$). If $(\Omega +\Omega _{c})>$ $\omega _{d}$ (the cut-off
frequency of the bath), which is the case in this figure, the function
$\mathfrak{S}(s,s-s^{\prime })$ oscillates faster than the rate of decay of the
bath $\Phi (s).$ The integral, $C(s)$,
of the product of the two functions generally oscillates around zero so
that $L(t)$ is reduced at any time $t$. The way to achieve this goal
(e.g., see Ref. \cite{Kurizki04}) is to
increase
the interaction of the pulse $\phi _{0}$ or decrease the pulse width $\tau $
in order to increase $\Omega _{c}=\phi _{0}/\tau .$

The dependence on the intensity is also of interest, as shown in
Fig. 1(C). The quality of control is seen to decrease with decreasing
intensity.

Finally, we consider the effects due to the different signs of $\phi _{0}$. The
two lower curves in Fig. 1(D) correspond to results for
different $\phi _{0}$ (or $\tau$)
but the same $\Omega _{c}=\phi _{0}/\tau $. The upper curves have the same
values but different signs. They show similar suppression, implying that
the quality of suppression depends primarily on $\Omega _{c}.$
Hence, the above discussions are also valid for the negative values of $\phi
_{0}$ as shown in the two upper curves.

For the case of a diatomic molecule, we note that the AC Stark effect
induced by an external laser field interacting with the electrons
decreases (or increases) $\Omega$, an effect termed
\textquotedblleft bond softening (or hardening)". \textit{Ab initio}
calculations \cite{Zavriyev90} show that, in the softening case, the
frequency can be reduced by ten percent for H$_{2}^{+}$ in a strong laser
field. In the case of hardening, if the frequency can be enhanced by  $p$
percent so that if $(1+p)\Omega >$ $\omega _{d}$, then leakage control
will be effective. This is expected to be a considerable technical challenge.

\textit{Conclusion.-} We have utilized an exact
trace-nonpreserving master equation for dynamics of a system
subspace, and introduced a leakage function that describes leakage
from a subspace of interest. We
used the second-order equation to analyze the quantum dynamics of
the leakage function, especially for the generic case of a
harmonic oscillator coupled to a bath of harmonic oscillators. The
realistic pulses required to suppress leakage in the presence of
the bath are given by the well
studied finite-width pulses causing AC Stark shifts
\cite{Kurizki04,Zavriyev90}.
A remarkable result is that, in the one-dimensional case, one can define a 
leakage function that provides a
complete description of quantum storage of a general superposition state
$|\phi\rangle$. Since quantum storage is of great
interest at present, the leakage function provides a most convenient tool
for studying the control of quantum states in the short time regime relevant to
quantum information processing and to coherent control.

\textit{Acknowledgement}: We thank Professor Daniel Lidar
for discussions with L.-A. Wu early in this work, and Goren Gordon and Noam Erez
for comments and discussions.
This work was supported by the NSERC Canada, a Varon Visiting Professorship
to P.B. from the Weizmann Institute, the ISF and EC (MIDAS and SCALA Projects).

\end{document}